# Conducting Atomic Force Microscopy Studies of Nanoscale Cobalt Silicide Schottky Barriers on Si(111) and Si(100)

J.L. Tedesco, J.E. Rowe
*Department of Physics, North Carolina State University, Raleigh, North Carolina 27695-8202, USA*

R.J. Nemanich
*Department of Physics, Arizona State University, Tempe, Arizona 85287-1504, USA*

## Abstract

Cobalt silicide ($CoSi_2$) islands have been formed by the deposition of thin films (~0.1 to 0.3 nm) of cobalt on clean Si(111) and Si(100) substrates in ultrahigh vacuum (UHV) followed by annealing to ~880ºC. Conducting atomic force microscopy has been performed on these islands to characterize and measure their current-voltage (I-V) characteristics. Current-voltage curves were analyzed using standard thermionic emission theory to obtain the Schottky barrier heights and ideality factors between the silicide islands and the silicon substrates. Current-voltage measurements were performed *ex situ* for one set of samples (termed "passivated surfaces") where the silicon surface surrounding the islands was passivated with a native oxide. Other samples (termed "clean surfaces") remained in UHV while I-V curves were recorded. By comparing the barrier heights and ideality factors for islands on passivated surfaces and clean surfaces, the effects of the non-passivated surfaces on conduction have been studied. The barrier heights measured from $CoSi_2$ islands on clean surfaces are found to be ~0.2 to 0.3 eV below barrier heights measured from similar islands on passivated surfaces. The main cause of the reduced Schottky barrier in the clean surface samples is attributed to Fermi level pinning by non-passivated surface states of the clean silicon surface. However, the measured barrier heights of the islands are equivalent on both clean Si(111) and Si(100) surfaces, suggesting that the non-passivated surface is influenced abstractby cobalt impurities. Furthermore, the barrier heights of islands on the clean surfaces are lower than can be explained by Fermi level pinning alone, suggesting the presence of additional reductions in the Schottky barrier heights. These variations are greater than can be attributed to experimental error, and the additional barrier height lowering is primarily attributed to spreading resistance effects. Schottky barrier inhomogeneity is also identified as a possible cause of the additional barrier height lowering and non-ideality in the Schottky contacts. Current-voltage measurements of the clean surface samples were also obtained at several temperatures. The barrier heights were found to decrease and the ideality factors were found to increase with decreasing temperature. The dependence of the barrier height is attributed to the temperature variation of the Fermi level.

**I. Introduction**

The Schottky barrier at the interface between a metal island and a semiconducting substrate will play an important role in the functionality of nanowire and nanodot electronic devices. It will allow nanowires to act as effective interconnects in nanoelectronic devices by electrically isolating them from the substrate [1]. The Schottky barrier has also been shown to be effective as a tunnel barrier in single electron tunneling devices [2]. Regardless of the specific means of implementation, a full understanding of the Schottky barrier is required for island-based devices.

One of the factors impeding such an understanding in these islands is that the barrier height is known to be related to the quality and morphology of the interface [3-4]. Cobalt silicide ($CoSi_2$) has a small lattice mismatch with silicon, -1.2% [5], which indicates that lattice-matched epitaxial layers can be formed on silicon [6-7]. However, the interfacial quality can influence the barrier height in a number of ways, ranging from barrier height inhomogeneity [8-9] to Fermi level pinning due to defects [10-12]. Furthermore, $CoSi_2$ has the added advantage that it is widely used in the semiconductor industry [13-15], offering the potential for a transition from the current device architecture to an island-based device architecture.

$CoSi_2$ islands have been grown on silicon substrates and studied using conducting atomic force microscopy (*c*-AFM) to record current-voltage (I-V) curves in both ultrahigh vacuum (UHV) and ambient conditions. Current-voltage curves have been recorded from the $CoSi_2$ islands both at and below room temperature. Using standard thermionic emission theory [16-17], the Schottky barrier heights, $\Phi_B$, and ideality factors, n, have been determined from the recorded I-V curves. By comparing barrier heights and ideality factors for both clean and passivated surfaces, as well as across multiple temperatures, conduction mechanisms across the nanoscale Schottky barrier have been investigated. Furthermore, the effects of the non-passivated surface on conduction have been studied.

**II. Experimental Methods**

Current-voltage curves were recorded from $CoSi_2$ islands on clean surfaces where the samples remained in UHV during the measurements. To study the differences in conduction through islands on clean and passivated surfaces, I-V measurements were also performed on $CoSi_2$ islands on passivated surfaces that were removed from UHV. Both types of samples were prepared using 25.4 mm diameter n-type silicon wafers with thicknesses of 0.25 to 0.28 ± 0.05 mm and phosphorous doping concentrations of ~$1 \times 10^{15}$ to $6 \times 10^{15}$ $cm^{-3}$. The doping concentrations were deduced from the resistivities, which were reported to range from 0.8 to 3.0 Ω-cm. The I-V measurements were performed using Veeco DDESP (doped diamond-like carbon coated silicon) cantilevers and Nanoworld CDT (doped diamond coated silicon) cantilevers. Both types of cantilevers were found to be equivalent for I-V measurements.

*II.A. Passivated Surfaces*

The silicon wafers were chemically cleaned *ex situ* using a combination of mercury lamp ultraviolet light-ozone (UV-ozone) and 10:1 hydrofluoric acid (HF) treatments. Cleaned wafers were mounted to molybdenum sample holders and secured with pieces of 0.125 mm diameter tantalum wire prior to being loaded into the UHV chamber for heat cleaning and deposition. The base pressure of the UHV chamber was $1.5 \times 10^{-10}$ torr.

Once under UHV conditions, the wafers were heated radiatively and held at a sample temperature of ~675ºC for 1 hour followed by heating to ~950ºC for 10 minutes. Sample temperatures were measured using an optical pyrometer. A Thermionics 150-0030 electron beam evaporation system was used to deposit 0.3 nm cobalt at room temperature at a rate of 0.02 nm/s. Following deposition, the sample was annealed at 880ºC for 25 minutes before cooling to room temperature. Following cool down, samples were removed from UHV, and the wafer backsides were swabbed with HF. Samples were then reintroduced into the UHV chamber where a layer of titanium ~100 to 200 nm thick was deposited at room temperature to form the backside ohmic contact.

Samples were removed from UHV and loaded into a commercially-available ambient AFM (ThermoMicroscopes Autoprobe CP-Research AFM) for topographical scanning and recording I-V measurements. I-V curves were recorded from 0 to 2.0 V in 0.02 V increments and only those curves where the current exceeded 10 µA at 2.0 V were processed. This was found to be the minimum level of conduction necessary to obtain consistency between measurements, and ~61% of the measurements met this condition. Whether those measurements that failed to meet this condition did so because of a highly resistive cantilever-island contact, irregularities within the island, or other reasons is unknown at this time. However, the contact resistance of the cantilevers used was deduced from I-V measurements on freshly-cleaved, highly oriented pyrolitic graphite (HOPG) using forces comparable to those used during the experiment. The contact resistance was measured to be several kΩ. The effect of this resistance was negligible at the low forward biases used in these measurements.

*II.B. Clean Surfaces*

The backsides of the silicon wafers were cleaned with a combination of UV-ozone and HF treatments. The wafers were loaded into the previously described UHV electron beam deposition chamber and a 200 nm thick cobalt layer was deposited on the backside surfaces.

Cobalt was used instead of titanium because a preliminary study indicated that a cobalt layer would be more likely than titanium to survive the high temperature flashing needed to clean the silicon surface in UHV without evaporating or diffusing into the silicon. Furthermore, as long as the cobalt layer was thick and the area covered it was large, it would still act as a sufficient backside ohmic contact. Following cobalt deposition, the wafers were removed and scribed into pieces ~2.5 × 10 mm$^2$, and single pieces were loaded into the UHV scanning probe system (Omicron Multiprobe P) without additional *ex situ* chemical cleaning. The UHV system consisted of a preparation chamber (base pressure: 7 × 10$^{-11}$ torr) and an analysis chamber (base pressure: 1.5 × 10$^{-11}$ torr). The preparation chamber was equipped with a triple-cell electron beam evaporator (EFM 3T). The analysis chamber was equipped with a variable temperature scanning tunneling microscope/atomic force microscope (Omicron VT AFM), and systems for Auger electron spectroscopy (AES) and low energy electron diffraction (LEED).

To limit outgassing during heat cleaning, the samples were radiatively heated for ~4 hours at a temperature of ~200 to 300ºC. Following the initial outgassing, samples were resistively heated and held at ~600ºC for at least 12 hours (typically overnight). Sample temperatures were measured using an optical pyrometer. Samples were then flashed using 3 to 4 second pulses at increasingly higher temperatures until the sample temperatures reached ~1125ºC. This temperature was maintained for 30 seconds 2 to 3 times before quickly cooling to 900ºC and then slowly cooling to room temperature.

Scanning tunneling microscopy (STM) was performed to ensure that the surface was nominally free of defects other than atomic steps. LEED and AES were performed to determine the long-range crystalline reconstruction of the surface, and to ensure that the surface was clean of contaminants, respectively. Once the surface was confirmed to be reasonably clean and ordered, ~0.1 to 0.2 nm of cobalt was deposited while the Si(111) samples were held at room temperature or while the Si(100) samples were held at 700ºC. The cobalt was deposited from pieces of 1.0 mm wire (Johnson Matthey, Grade 1 purity) in a molybdenum crucible. Following deposition, AES was performed to verify the cobalt deposition and that the surface did not show a significant increase in contamination. The sample was then annealed to ~880ºC for 15 to 30 minutes. After annealing, LEED patterns were obtained, and the presence of a Si(111):7×7 or Si(100):2×1 diffraction pattern indicated the formation of $CoSi_2$ islands [18].

Samples were then transferred to the microscope stage for *c*-AFM measurements. I-V curves were recorded at room temperature first to locate and select islands that demonstrated high conduction at low voltages (I > 333 nA at V < 0.5 V) and rectifying behavior at reverse biases (I ~ 10 nA at V > 1.0 V). Once such islands were identified, the sample temperature was slowly lowered and I-V curves were recorded on the identified islands at several temperatures between room temperature and 75 K.

**III. Results**

*III.A. $CoSi_2$ Island Topography*

For all surfaces, the $CoSi_2$ films formed into faceted and presumably epitaxial islands upon annealing to ~880ºC. As shown in Figs. 1(a) and 1(b), triangular and non-triangular islands formed on the Si(111) surfaces. The density of islands is lower in Fig. 1(b) presumably because the original $CoSi_2$ film was thinner. The $CoSi_2$ islands on Si(100) typically formed into long rectangular islands oriented along orthogonal <110> directions, as shown in Fig. 1(c).

*III.B. Electrical Characteristics of CoSi$_2$ Islands*

Room temperature Schottky barrier height values determined from I-V curves for islands on passivated Si(111) surfaces were found to range from 0.53 to 0.63 ± 0.02 eV, with two outliers at ~0.45 eV, and the values displayed a linear correlation to the ideality factors. For the room temperature measurements of islands on clean Si(111) the values of the barrier heights ranged from 0.30 to 0.45 ± 0.01 eV and values of the ideality factors ranged from 1.09 to 1.83 ± 0.05. A correlation between the barrier height and the ideality factor was not evident. Room temperature values of the Schottky barrier height for islands on clean Si(100) range from 0.39 to 0.46 ± 0.02 eV with values of the ideality factor ranging from 1.06 to 1.40 ± 0.05. Again, a correlation between the barrier height and ideality factor was not evident. The errors stated above are a combination of the experimental and analytical uncertainties associated with those measurements. The experimental uncertainty for the measurements from each island was individually determined by comparing the maximum, minimum, and average values for barrier heights and ideality factors recorded from that island. The analytical uncertainties for the Schottky barrier heights, $\Phi_B$, and ideality factors, n, were determined by calculating the total differentials of $\Phi_B$ and n using the thermionic emission equations. First, the equations for $\Phi_B$ and n were partially differentiated in terms of the variables in the equations (i.e. $\partial \Phi_B / \partial T$). Each derivative was then multiplied by an average experimental uncertainty associated with the variable of the partial derivative (i.e. $(\partial \Phi_B / \partial T) \cdot \Delta T$). The products were then summed separately for the barrier height and the ideality factor. The sums were taken as the analytical uncertainties for the two quantities and were assumed to be constant for all measurements.

The values of barrier heights and ideality factors for all three sets of samples are shown graphically in Fig. 2. The dashed line through the data from the islands on passivated Si(111) is a linear fit to the data points shown (which excludes the outliers at ~0.45 eV). The dotted lines through the data points from the islands on clean Si(111) and clean Si(100) are also linear fits to those data points. Both fits to the clean surface data points, however, are significantly weaker than the fit to the passivated surface data. The relationships between barrier height and island area for all three sets of islands are shown in Fig. 3. The barrier heights for all three sets of samples tend to decrease with decreasing area, and this trend is most evident for the islands on clean Si(111) with areas below 100,000 nm$^2$. The relationship between barrier height and the ratio of island area-to-island periphery is shown in Fig. 4. A decreasing ratio of island area-to island periphery is generally indicative of decreasing island size. As shown in Fig. 4, the barrier height decreases with decreasing ratio for the islands on clean Si(111). However, a correlation between decreasing ratio and barrier height was not evident for the islands on passivated Si(111) and clean Si(100).

Temperature-dependent I-V data was collected only for the clean surface samples. Plots of barrier height as a function of temperature for four islands grown on clean Si(111) and two islands grown on clean Si(100) are shown in Figs. 5(a) and 5(b), respectively. The Schottky barrier heights for the islands on Si(111) decrease with decreasing temperature from ~0.38 eV to ~0.21 eV, depending on the island. Schottky barrier heights for islands on Si(100) exhibited similar trends, decreasing from ~0.41 eV to ~0.18 eV, depending on the island. The ideality factors for the islands on Si(111) increased from ~1.13 to ~2.16 over the same temperature range, depending on the island. The ideality factors for the islands on Si(100) increased from ~1.08 to ~3.03, depending on the island. All islands in this study exhibited these same general trends of decreasing Schottky barrier height and increasing ideality factor with decreasing temperature.

**IV. Discussion**

*IV.A. CoSi$_2$ Island Topography*

The fact that triangular islands grow on the Si(111) surface is expected due to the three-fold symmetry of the underlying substrate [19]. Some islands exhibit more complex structures but with edges still oriented along the triangular directions. Triangular and partially triangular islands in close proximity to each other have been reported previously [6,19-20] and these partially triangular shapes suggest that the shape is evolving toward triangular. A similar growth process occurs for epitaxial islands of DySi$_2$ on Si(111), in which irregular islands grow into stable faceted triangular islands [21]. The triangular islands are known to have two orientations based on the epitaxial relationships between the island and the substrate [19-20]. It was observed that in some regions, most triangular islands had the same orientation, suggesting a similar epitaxial relationship, but this could not be verified with these measurements. Furthermore, it was not possible to determine the orientation of the non-triangular islands.

Rectangular CoSi$_2$ islands on Si(100) oriented along orthogonal <110> directions have also been reported previously [22]. While neither cross-sectional transmission electron microscopy nor selected area electron diffraction measurements were performed, the growth conditions for the current study and the previous study [22] were similar. Therefore, it is likely that the epitaxial relationship between the rectangular CoSi$_2$ islands and the Si(100) surface is the same for both studies. The surfaces of the rectangular islands grow into the surface along the {111} and {511} planes and two orientations are possible along each of the

<110> directions. The two orientations are rotated 70.54° relative to each other, but otherwise, the structures of the differently-oriented islands are the same [22].

*IV.B. Electrical Characteristics of $CoSi_2$ Islands*

The relatively large ideality factors suggest complex processes at the interface [4]. A previous study [23] proposed that the non-ideality of the interface is related to a bias-dependent barrier height. The mechanisms potentially responsible for a bias-dependent barrier height include image force lowering, interface states, hole injection, carrier recombination in the depletion region, and thermionic field emission due to field enhancement [23-25].

The doping concentrations of the substrates are ~$10^{15}$ cm$^{-3}$. The dominant means of current transport in a Schottky contact is determined by the relation between kT and $E_{00}$ [16-17]. The value of $E_{00}$ is a measure of the importance of tunneling through a Schottky barrier from a semiconductor into a metal at a given temperature [16]. For barriers on n-type semiconductors, $E_{00} = (e\hbar/2)(N_D/\varepsilon_s m^*)^{1/2}$, where $e$ is the charge of the electron, $N_D$ is the donor concentration of the semiconductor, $\varepsilon_s$ is the permittivity of the semiconductor, and $m^*$ is effective mass of the electron [17]. When $E_{00} \ll kT$, thermionic emission is the dominant form of current transport [16-17]. For the samples in this study, $E_{00}$ is ~0.3 to 0.8 meV [16], which is significantly less than kT at any temperature in this study. Therefore, the current transport in each sample is in the thermionic emission regime [17], which is in agreement with previous studies [24-25]. Therefore, effects due to thermionic field emission or field emission should not be significant at any temperature range in this study [26-28]. Additionally, a previous study has found that when $N_D$ ~ $10^{15}$ cm$^{-3}$, the reduction in the barrier height due to image force lowering is less than 0.01 eV [29]. Thus, image force lowering is a negligible effect at the low doping concentrations used in this study. Previous studies performed using similar substrates have reached the same conclusions [24-25]. To investigate possible effects due to field enhancement, the values of kT vs. nkT for the islands are plotted to establish whether thermionic field emission occurs for the temperature range studied. If thermionic field emission is important, then the fit of the kT vs. nkT plot would become non-linear and approach a constant value at low temperature [30-31]. For this study, the fit is linear throughout the temperature range studied; suggesting that thermionic field emission due to field enhancement is not the dominant effect.

The correlation between barrier height and ideality factor exhibited by the passivated samples has been noted in previous studies of Schottky contacts [4,25,32-36]. Furthermore, the lack of correlation exhibited by the clean surface samples has also been demonstrated in a previous study [11]. The correlation between barrier height and ideality factor for islands on passivated samples suggests that the reduced barrier height is related to the quality of the interface [4]. However, the lack of a correlation exhibited by islands on the clean surfaces indicates that there is another mechanism responsible for the barrier height lowering in these samples.

The range of barrier heights measured from islands on clean surfaces is ~0.2 to 0.3 eV below typical values reported in the literature for macroscopic $CoSi_2$/n-Si contacts [5,37-39] and is centered around ~0.4 eV, as shown in Fig. 2. This shift may be attributed to surface states of the clean surface that pin the Fermi level of the silicon surface on the perimeter of the islands. Previous photoemission studies have found that the Fermi level of the clean Si(111) surface is pinned relative to the valence band maximum at either $0.63 \pm 0.05$ eV [40] or 0.55 eV [41]. Another photoemission study reported that the Fermi level of the clean

Si(100) surface is pinned relative to the valence band maximum at ~0.4 eV [42]. Possible band diagrams of the interface are shown in Fig. 6, which display how the Fermi level pinning at the island periphery could lead to a negative shift in the barrier height. The surface states create a surface dipole energy $\Delta$, which is shown schematically in Fig. 6(b). Gap states at the interface between the island and the silicon substrate should be present at the interfaces of islands on both passivated and clean surfaces. However, because the islands cover only 10 to 20% of the surface, the pinning states on the clean surfaces surrounding the islands will impact the measured barrier height more than the interfacial gap states.

It must be noted, however, that the measured barrier heights in Fig. 2 are similar in magnitude for islands on both clean Si(111) and clean Si(100) surfaces. Measured barrier heights of islands on the clean Si(111) and clean Si(100) surfaces should differ by ~0.15 to 0.2 eV between the two surfaces [40-42]. Thus, the similarity in measured barrier heights shown in Fig. 2 suggests that the "clean" surfaces in this study are influenced by the presence of cobalt impurities from the deposition. Nevertheless, the Schottky barrier lowering shown in Fig. 2 is suggestive of Fermi level pinning by surface states of the clean silicon surface surrounding the $CoSi_2$ islands.

However, the distribution of barrier heights shown in Figs. 2 and 3 suggest that there is additional barrier height lowering beyond that which is expected due to Fermi level pinning alone. The influence of spreading resistance must be considered as a possible cause for the additional barrier height lowering in these samples. In circular contacts, the spreading resistance is inversely proportional to the radius of the contact [43-44]. Though the smaller islands are not circular, they will most likely experience a significantly increased spreading resistance relative to that of the larger islands, leading to lower measured barrier heights.

Therefore, increased spreading resistance would create an apparent trend toward lower barrier heights with decreasing island area, as shown in Fig. 3. Additionally, as shown in Fig. 4, the barrier height decreases with the ratio of the island area-to-island periphery. Decreasing barrier heights with decreasing island area-to-island periphery ratio have been reported to be due to increases in recombination in small islands due to increased electric fields at their edges [24]. However, as shown in Fig. 4, the pronounced decrease in barrier height with ratio is evident only for islands on the clean surfaces. Furthermore, as stated previously, according to plots of kT vs. nkT, there is no evidence of field enhancement in islands on the clean surfaces. Therefore, it is unlikely that recombination is a significant effect in the additional barrier height lowering shown in Figs. 2 and 3.

Figure 5 indicates that as the temperature decreases, the barrier height decreases. This effect may be due to the temperature dependence of the Fermi level. It can be calculated that for samples with $N_D \sim 10^{15}$ cm$^{-3}$, the barrier height will decrease by 0.20 to 0.25 eV due to the temperature dependence of the Fermi level [45]. A decrease of 0.20 to 0.25 eV would be of a similar magnitude to the decrease in the barrier height shown in Fig. 5. Similar temperature-dependent phenomena [5] have also been associated with Schottky barrier height inhomogeneity [8]. According to the barrier height inhomogeneity model, the island-substrate interface may be composed of several regions with differing barrier heights. As the temperature decreases, regions with low barrier heights begin to dominate conduction [46]. Thus, the barrier height decreases with temperature. However, this effect would only impact the measured barrier height if the region was significantly larger than the "critical area" [46]. For $N_D \sim 10^{15}$ cm$^{-3}$, the critical area is ~50,000 nm$^2$ [46]. While the areas of most of the islands exceed this value, some do not, as shown in Fig. 3. However, these smallest islands still exhibit significant barrier lowering. Therefore, while Schottky barrier inhomogeneity may be present, it is most likely a minor effect.

Further experiments would be necessary to determine the full impact of pinning by surface states, barrier height inhomogeneity, and other defects. Passivation studies performed entirely in UHV could be utilized to determine the extent to which surface states lead to Schottky barrier lowering. Ballistic electron emission microscopy studies could be used to image the interface to determine if the interfaces are composed of patches of different barrier heights. These additional studies are beyond the scope of the work presented here. However, they would provide information important in the implementation of defect engineering schemes to take advantage of these Schottky barrier lowering effects. It is also important to note that the defect engineering schemes that utilize pinning by surface states would have to be designed so that the surface states survived the process sequence.

**V. Summary and Conclusions**

Nanoscale $CoSi_2$ islands were grown on Si(100) and Si(111) substrates and I-V measurements were performed using *c*-AFM, both at and below room temperature. In these measurements, two different values of the Schottky barrier height were found depending on the sample preparation procedure. In "passivated surface" island samples, the range of barrier heights measured on the $CoSi_2$ islands approached the range of barrier heights typically reported in the literature. Furthermore, a linear correlation was found between decreasing Schottky barrier height and increasing ideality factor, as well as a correlation between increasing barrier height and increasing island area. No correlation was found between barrier height and ideality factor for the measurements recorded from $CoSi_2$ islands on "clean surface" samples at room temperature. However, a correlation between decreasing barrier height and decreasing island area was found. For the "clean surface" island samples, the range of barrier heights measured was ~0.2 to 0.3 eV below the range reported in the literature for these contacts. This shift in the barrier heights is attributed to Fermi level pinning by the non-passivated surface states of the clean silicon surface surrounding the $CoSi_2$ islands. Additionally, the fact that the barrier heights measured from islands on both substrates are similar suggests that the surface was influenced by the presence of cobalt impurities. Furthermore, there was additional barrier height lowering beyond what was ascribed to Fermi level pinning alone. The correlations between barrier height and island area suggested that the primary source of the additional barrier height lowering and non-ideal behavior was spreading resistance effects. Barrier height inhomogeneity was also considered as a reason for barrier height lowering and non-ideality, but it was concluded to be a minor effect. Using I-V-T measurements, the temperature-dependent electrical characteristics of these islands were investigated. Schottky barrier heights were found to decrease with decreasing temperature while the ideality factors increased with decreasing temperature. The temperature-dependent behavior of Schottky barrier heights was attributed to the temperature dependence of the Fermi level. In summary, the evidence in this study suggests that surface states of the non-passivated silicon surface are critical in determining the measured Schottky barrier heights of $CoSi_2$ islands. Furthermore, by making controlled use of these surface states (e.g. defect engineering) it should be possible to tune the barrier height without changing the stable chemical composition of the surface. Further passivation studies would be necessary to test this tuning effect; however, it could become significant in device design and manufacturing using silicide island metal contacts.

**Acknowledgements**

This work was supported by the National Science Foundation through Grant No. DMR-0512591.

**Figures**

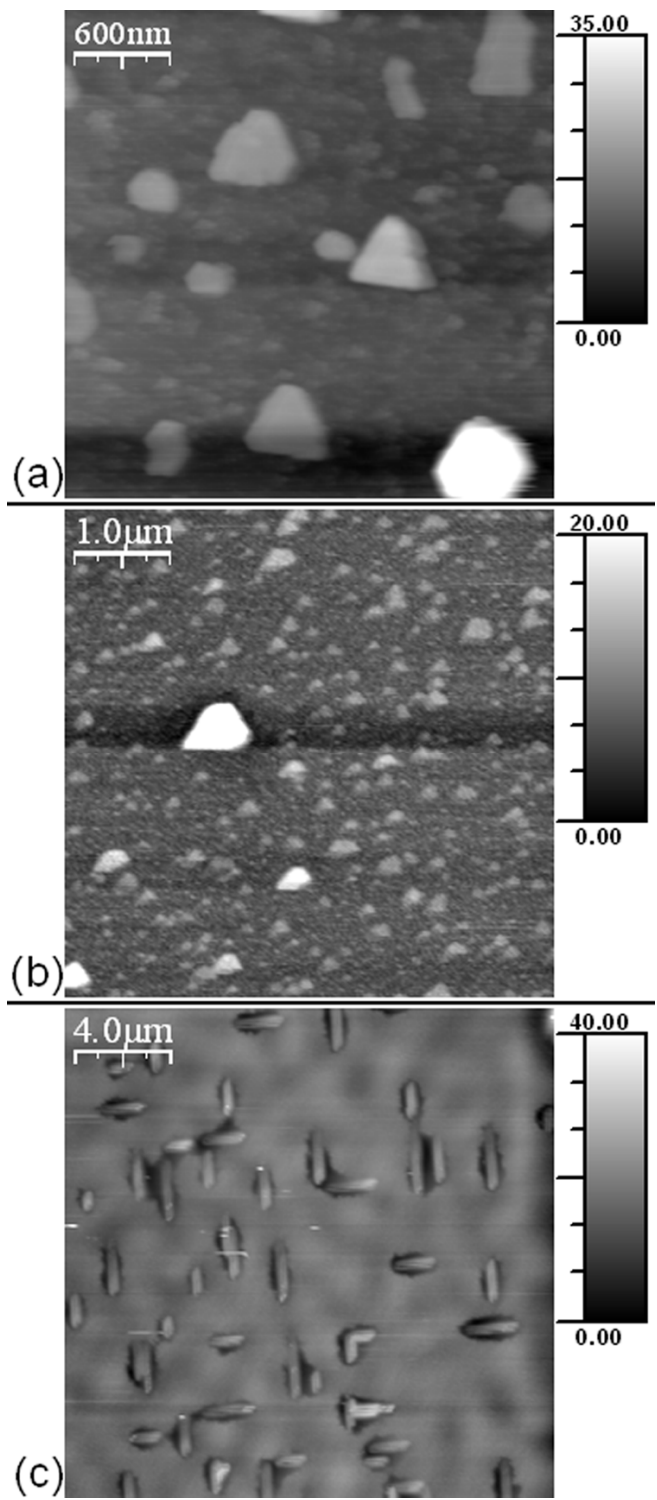

Fig. 1. AFM topography images of CoSi$_2$ islands on (a) the passivated Si(111) surface (scan size: 3 μm), (b) the clean Si(111) surface (scan size: 5 μm), and (c) the clean Si(100) surface (scan size: 20 μm). The height scale bars are in units of nanometers.

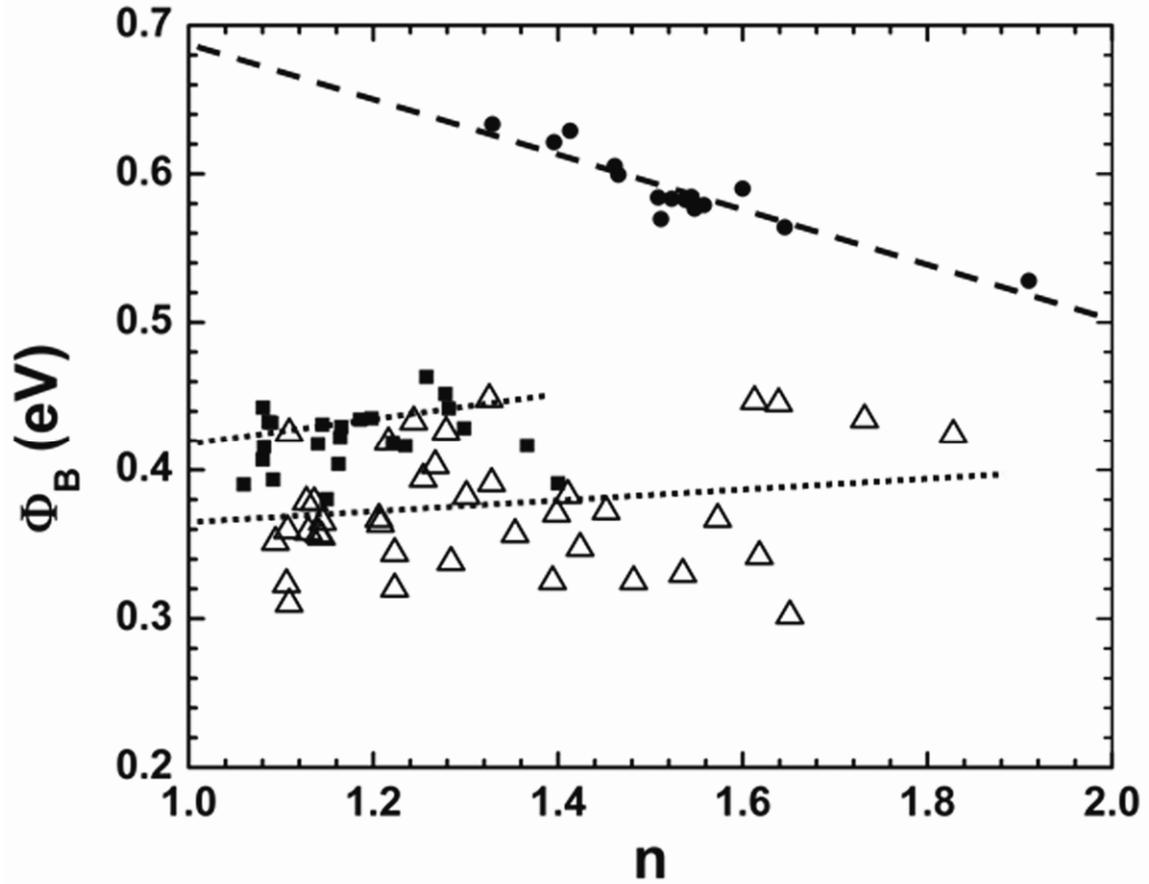

Fig. 2. Schottky barrier heights and ideality factors for islands on passivated Si(111) surfaces (●), islands on clean Si(111) surfaces (△), and islands on clean Si(100) surfaces (■). The dashed line is a linear fit through the data from islands on passivated Si(111), while the dotted lines are linear fits through the data from islands on clean Si(111) and Si(100) surfaces.

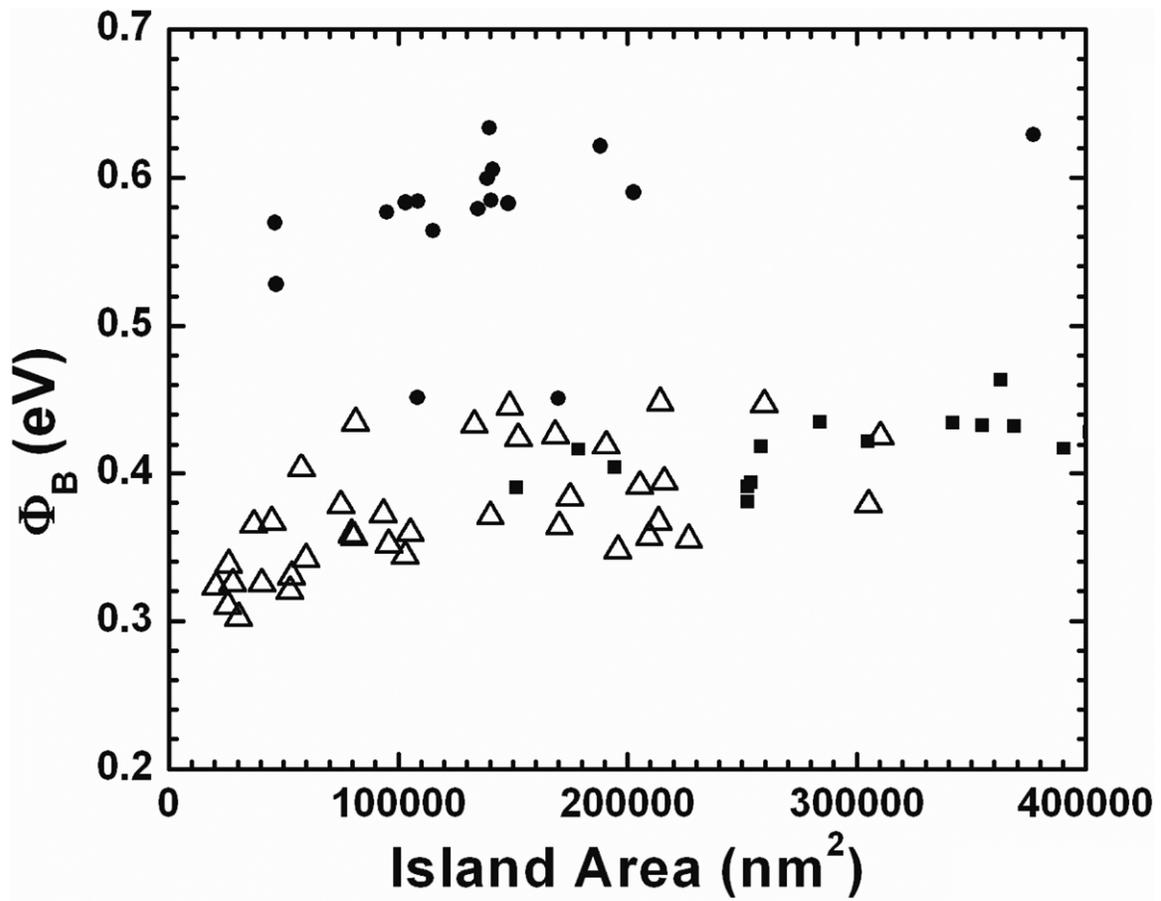

Fig. 3. Room temperature areal relationships for the Schottky barrier heights of islands on passivated Si(111) surfaces (●), islands on clean Si(111) surfaces (△), and islands on clean Si(100) surfaces (■).

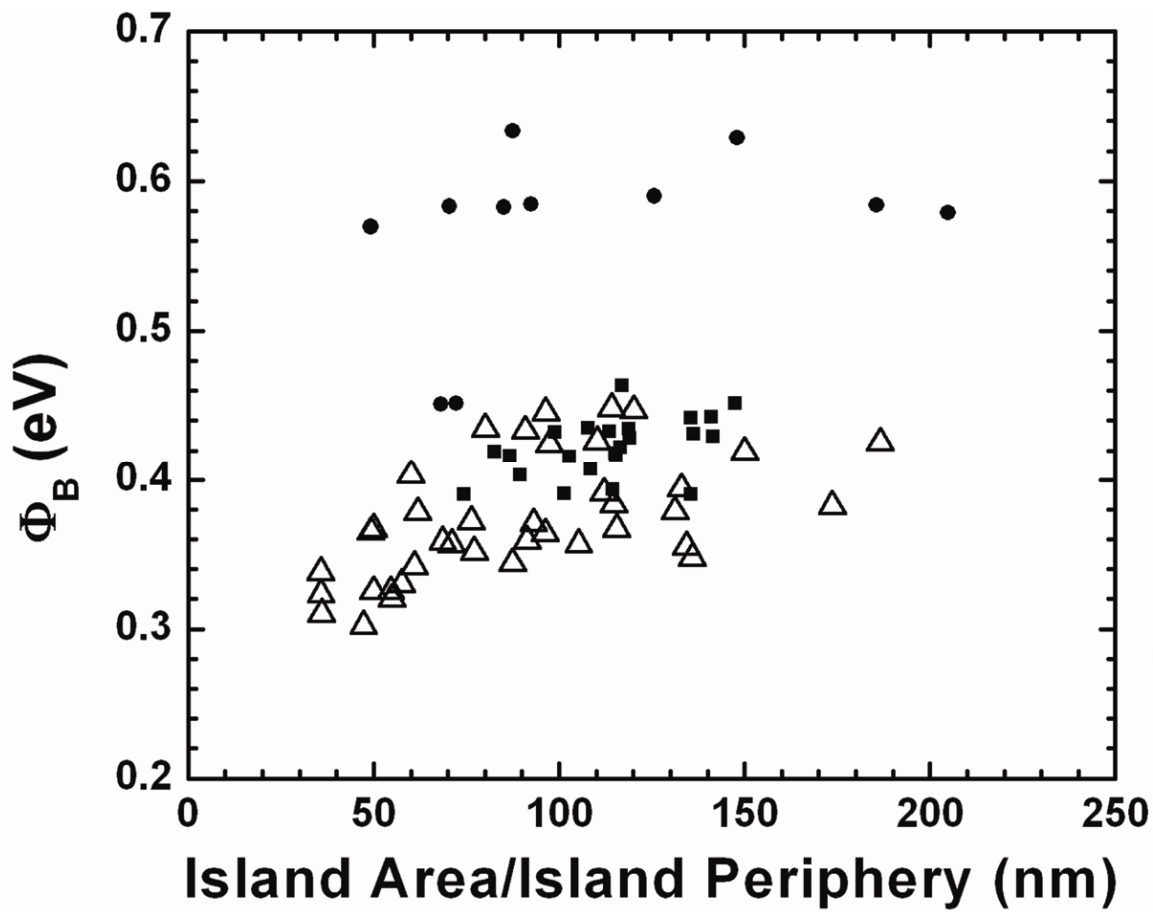

Fig. 4. Room temperature relationships between the Schottky barrier heights and the island area-to-island periphery ratios for $CoSi_2$ islands. The plot includes islands on passivated Si(111) surfaces (●), islands on clean Si(111) surfaces (△), and islands on clean Si(100) surfaces (■).

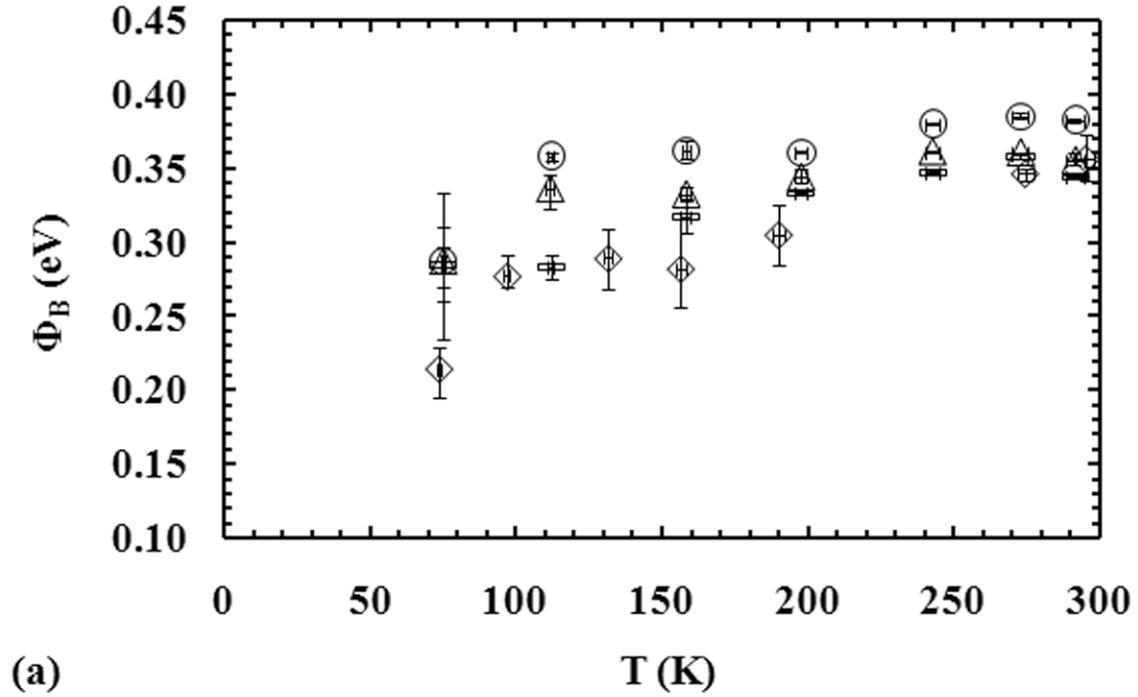

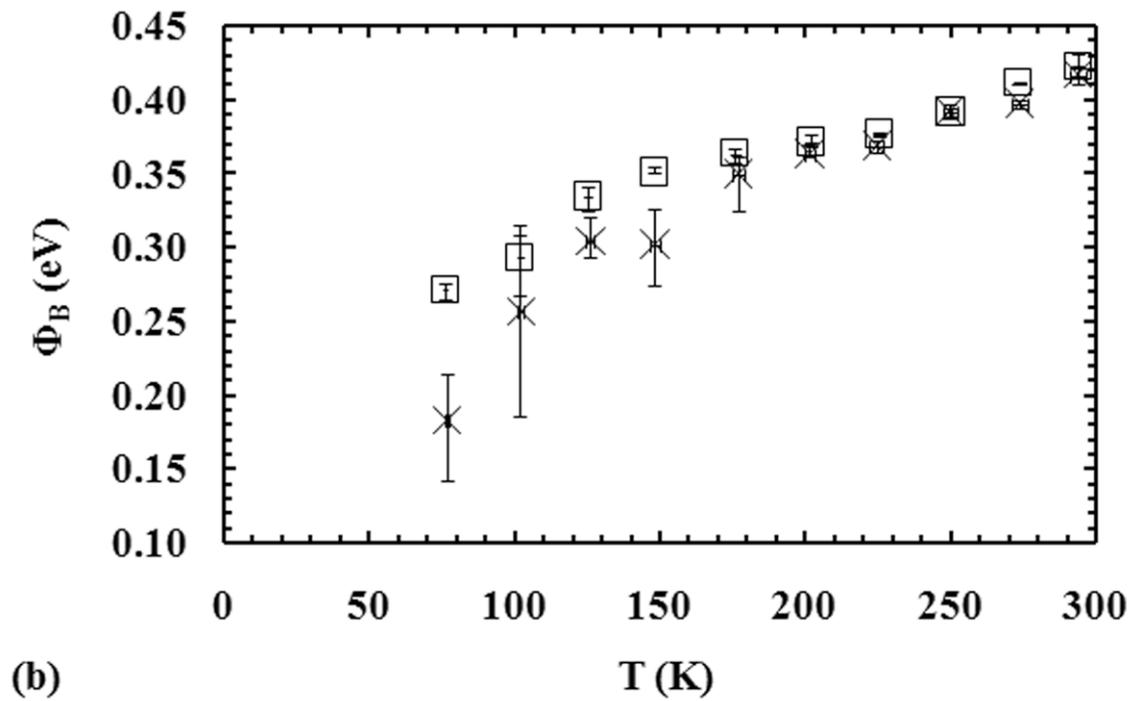

Fig. 5. Schottky barrier heights as a function of temperature for $CoSi_2$ islands on (a) the clean Si(111) surface and (b) the clean Si(100) surface. The different symbols are only present to differentiate between the data from different islands and have no intrinsic significance.

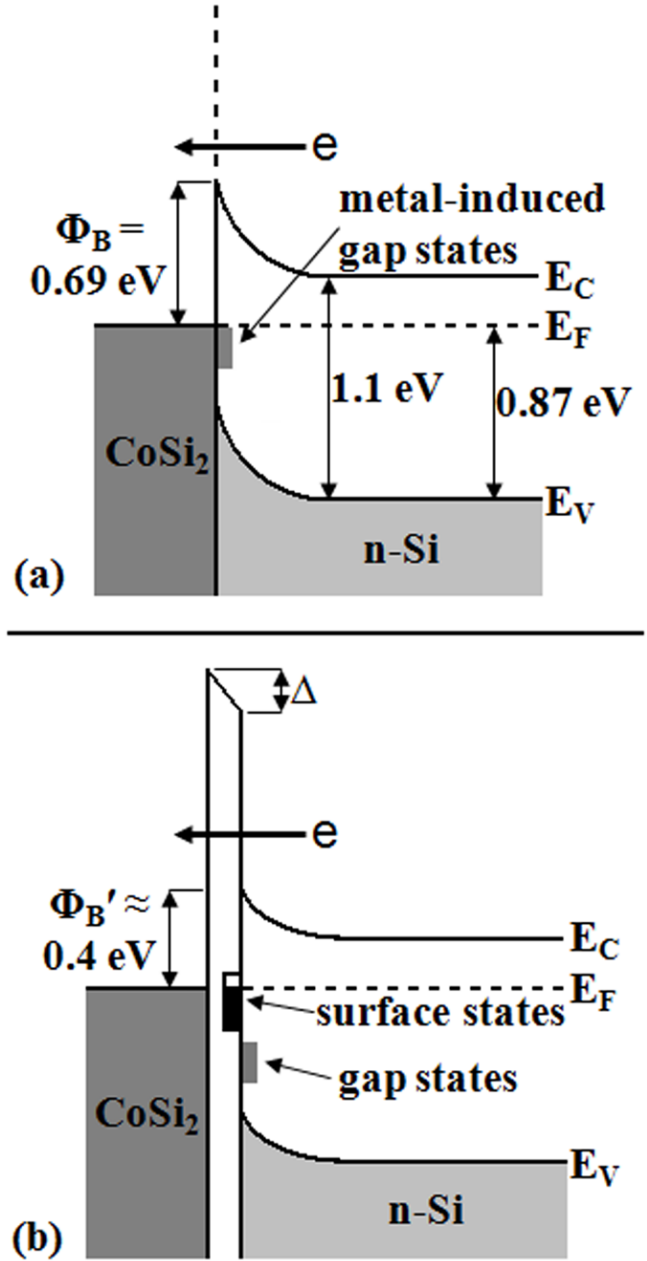

Fig. 6. (a) Band diagram of the passivated CoSi$_2$/n-Si interface. The bold arrow labeled "e" indicates the direction of current flow in forward bias. (b) Band diagram of the clean CoSi$_2$/n-Si interface at the periphery of the CoSi$_2$ islands demonstrating how Fermi level pinning lowers the Schottky barrier height. $\Delta$ is the peripheral dipole energy and $\Phi_B' = \Phi_B - \Delta$. Note that the band bending is reduced by ~0.25 to 0.3 eV due to the interface dipole barrier. The metal-induced gap states are present but do not affect the measured barrier height due to pinning by peripheral surface states.